\documentclass{PoS}

\usepackage{subcaption}
\usepackage{textcomp}

\title{Novel Back-coated Glass Mirrors for the MAGIC Telescopes}

\ShortTitle{Novel Back-coated Glass Mirrors}

\author{\speaker{Martin Will}\\
        Max-Planck-Institut f\"ur Physik, Munich, Germany\\
        E-mail: \email{mwill@mpp.mpg.de}}

\author{Juliane van Scherpenberg\\
        Max-Planck-Institut f\"ur Physik, Munich, Germany\\
        E-mail: \email{jvsch@mpp.mpg.de}}

\author{Razmik Mirzoyan\\
        Max-Planck-Institut f\"ur Physik, Munich, Germany\\
        E-mail: \email{razmik.mirzoyan@mpp.mpg.de}}

\author{Markus Garczarczyk\\
        Deutsches Elektronen-Synchrotron (DESY), Zeuthen, Germany\\
        E-mail: \email{markus.garczarczyk@desy.de}}

\author{for the MAGIC Collaboration\footnote{\texttt{https://magic.mpp.mpg.de}. For collaboration list see PoS(ICRC2019)1177}}

\abstract{
  The mirrors installed on Imaging Atmospheric Cherenkov Telescopes like the
  MAGIC telescopes in La Palma, Canary Islands, are constantly exposed to the
  harsh environment. They have to withstand wind-induced corrosion from dust and
  sand, changing temperatures, and rain. Because of the size of the telescope,
  protecting the structure with a dome is not practical. The current mirrors
  used in MAGIC are aluminum front-coated glass mirrors, covered by a thin
  quartz layer. But even with this protective layer, significant decrease in
  reflectivity can be seen on timescales of several years. The quartz layer is
  very delicate and can be easily scratched or damaged, which also makes
  cleaning the mirrors almost impossible. We have tested a novel design of glass
  mirrors that can be easily cleaned and should show almost no degradation in
  reflectivity due to environmental influences. The protective layer is a
  ultra-thin glass sheet which is back-coated with aluminum, making it possible
  to simply wipe the mirror with household cleaning tools. In this contribution
  we will present results from laboratory tests of reflectivity and focusing
  properties of prototype mirrors, as well as long-term tests on-site at the
  MAGIC telescopes. We will also outline plans for exchanging a large fraction
  of MAGIC mirrors with this novel design, guaranteeing a peak performance of
  MAGIC for the coming years.
}

\FullConference{36th International Cosmic Ray Conference -ICRC2019-\\
		July 24th - August 1st, 2019\\
		Madison, WI, U.S.A.}

\begin{document}

\section{Introduction}

Current Imaging Atmospheric Cherenkov Telescopes (IACT) need large reflective
surfaces, about 500\,m$°2$ in the case of
MAGIC\footnote{\href{https://magic.mpp.mpg.de}{magic.mpp.mpg.de}}. Future IACTs
will be even larger, several thousand m$^2$ for the Cherenkov Telescope Array
(CTA)\footnote{\href{https://www.cta-observatory.org}{cta-observatory.org}}.
This makes it very costly and practically impossible to protect the telescope
with a dome, they are constantly exposed to the environment with the need for
regular maintenance and facing significant degradation over timescales of a few
years. A serious effort went into the development of a new mirror design, which
could be easier to maintain and have a longer. To be installed in MAGIC they
have to meet stringent requirements:
\begin{itemize}
  \item \textbf{High reflectivity:}
    The reflectivity of the mirrors should be very high ($>$70\%), especially in
    the wavelength range where the Cherenkov spectrum peaks ($\sim$ 350\,nm) and
    the photo sensors in the camera are most sensitive.
  \item \textbf{Good focusing properties:}
    The collected light must be well focused onto the camera. Ideally, the
    reflected spot of a point-like light source at the focal plane of the mirror
    is smaller than the area of one pixel in the camera.
  \item \textbf{High resistance against environmental influences:}
    The mirrors are constantly exposed to harsh weather conditions at desert and
    high-mountain altitudes, they must show high resistance to UV light, rain,
    ice, high wind, and corrosion from sand.
  \item \textbf{Different curvature radii:}
    For the overall parabolic shape, the spherical mirrors must have specific
    radii (34.5--36\,m in MAGIC) depending on their location in the dish.
  \item \textbf{High stiffness:}
    The mirrors must be rigid enough to avoid deformation due to varying
    gravitational loads, which could result in additional image aberrations.
  \item \textbf{Low weight:}
    At the same time the mirrors should be as light-weight as possible to
    minimize the load on the drive system and allow for faster movement of the
    telescope.
  \item \textbf{Low cost:}
    In order to produce the quantity needed to furnish both MAGIC telescopes
    (almost 500 mirrors with a surface of 1\,m$^2$ each), the production process
    needs to be time and cost-effective.
\end{itemize}

\section{Mirror Design for Cherenkov Telescopes}

The mirrors installed in MAGIC can be divided in two categories, both having a
light-weight aluminum honeycomb plate of 2--6~cm thickness in the structure for
stiffness and stability. The reflective surface is either diamond-milled
aluminum or a cold-slumped glass sheet covered with an aluminum layer. Both
types of mirrors are shaped with the desired curvature.

To protect the reflective aluminum layer from oxidation, a 100\,nm thin quartz
layer is vacuum-deposited on the surface. Quartz is well suited for this since
it is a very robust material and hardly reacts chemically. The thickness of the
quartz layer was optimized to enhance the reflectivity at the peak wavelength of
the Cherenkov spectrum through positive interference, trading durability for
reflectivity. The quartz layer can prevent dust particles from scratching the
aluminum layer and chemicals, which are contained in the air and rain, from
reacting with the aluminum and altering its reflective
properties~\cite{reflectivesurface}. However, there are limitations to this
approach, the surface becomes very delicate and the lifetime of the mirror is
limited since the protective layer is not perfect. Possible microscopic cracks
in the quartz layer, and the wearing due to the day and night strong temperature
variations, could allow acids in rain water to enter through the protective
layer, reach the aluminum and oxidize or alter the chemical composition of the
aluminum. Over time, this can lead to a significant loss of reflectivity.

Another effect, which mostly produces a short term loss in reflectivity and
affects both mirror types equally, is the deposition of dust and dirt. Since
MAGIC is located in arid environments, it is inevitable that over time sand will
deposit on the reflector, decreasing the reflectivity. It is not trivial to
remove the dust from the mirrors, because the thin layers of quartz and aluminum
are delicate and have to be treated with great care, and hence active cleaning
is not possible. Rain water can remove accumulated dust, but oily residues stick
to the surface and cannot be removed. Other IACT facilities regularly re-coat
the mirrors or perform major refurbishment~\cite{VeritasCleaning,
HESSrecoating}. In MAGIC, the performance of each single mirror is regularly
evaluated, and those showing a strong degradation are
exchanged~\cite{absolutereflectance}. All of these procedures are expensive and
time consuming, the merit of developing new strategies for prolonging the
lifetime and high-reflectivity of the mirrors.

\section{Novel Back-Coated Glass Mirrors}

A new design is being developed by MAGIC scientists together with the Italian
mirror manufacturer Media Lario Technologies
(MLT)\footnote{\href{https://www.medialario.com}{www.medialario.com}}. It uses a
thin glass sheet as cover of the reflective layer, that is deposited on its rear
side. This addresses the two major sources of reflectivity loss: the deposition
of dust and other residues, and the degradation due to acid rain chemically
reacting with the aluminum. The glass sheet can be easily cleaned without risk
of damaging the mirror, and glass is very resistant to any chemicals and acid
cannot enter. The new mirrors should not lose transmissivity over decades of
exposure to strong weather.

\begin{figure}[ht]
  \centering
  \begin{subfigure}{.535\textwidth}
    \centering
    \includegraphics[width=\columnwidth]{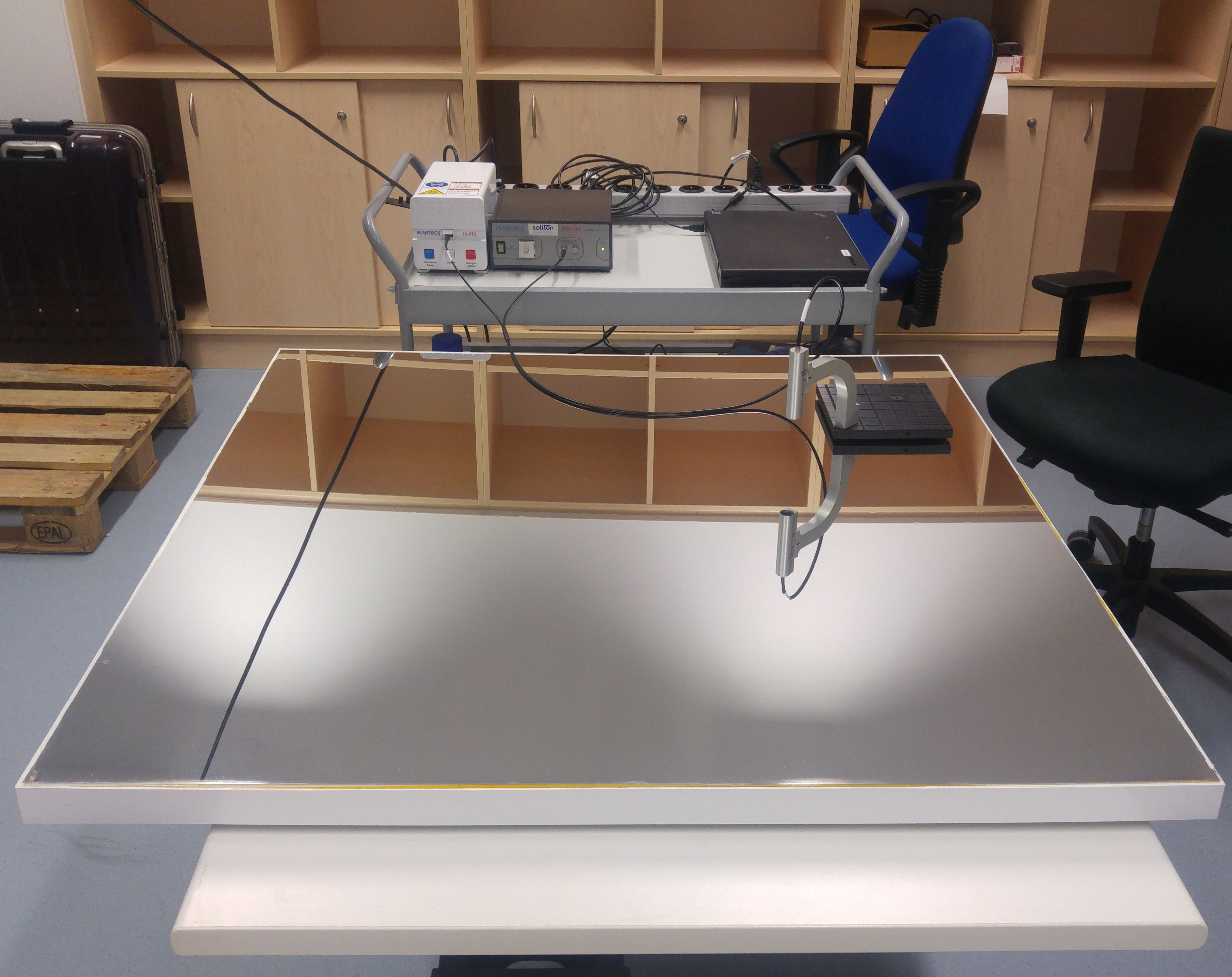}
  \end{subfigure}
  \hspace{1em}
  \begin{subfigure}{.23\textwidth}
    \centering
    \begin{subfigure}{\textwidth}
      \centering
      \includegraphics[width=\columnwidth]{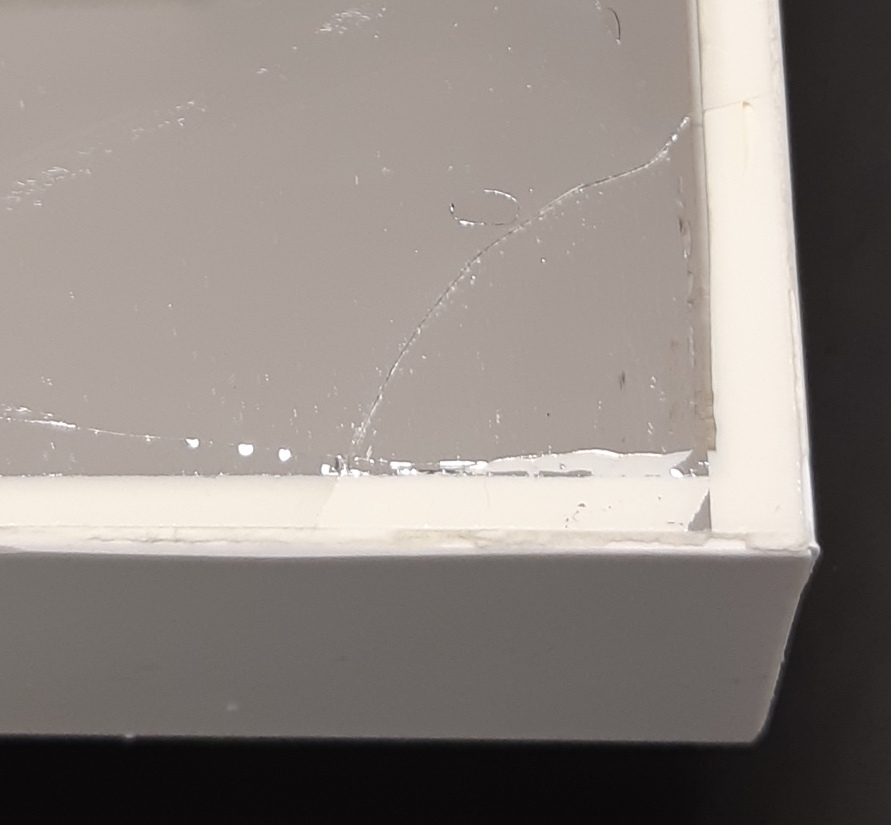}
    \end{subfigure}
    \begin{subfigure}{\textwidth}
      \centering
      \includegraphics[width=\columnwidth]{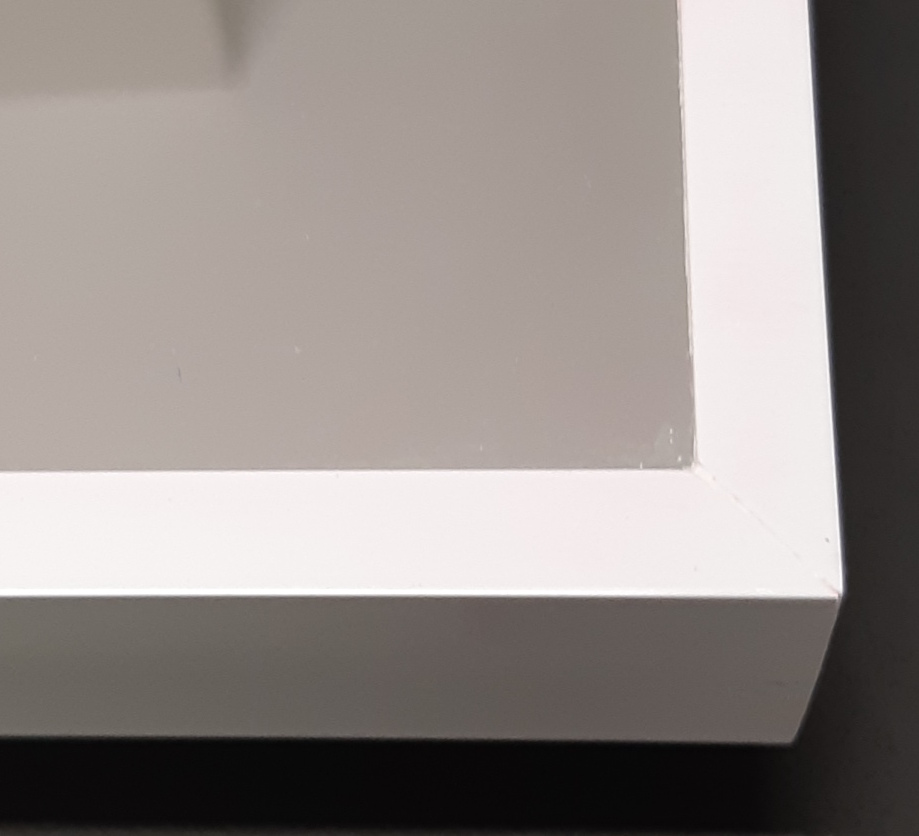}
    \end{subfigure}
  \end{subfigure}
  \caption{Left: Laboratory setup at MPP to measure the surface reflectivity.
  Right: Closeup of the corners of two mirrors (top: ID~224, bottom: ID~229)
  from different production periods. On the top, the thin glass sheet is not
  bonded well with the backing structure and air bubbles are seen, as well as
  cracks close to the edge. On the bottom, one of the latest mirrors is seen
  with edge protection to prevent cracking, better glue distribution and fewer
  defects are seen.}
  \label{fig:edge_closeup}
\end{figure}

This is achieved by adapting the production process of cold-slumping glass
mirrors~\cite{coldslumping, coldshaping}. One side of a 0.4\,mm thin glass sheet
is covered with aluminum and glued to the sandwich structure, the uncoated side
facing up. A lot of care has been taken to not damage the thin glass sheet in
the gluing process, especially at the edges of the sheet. The thin glass sheet
is slightly larger than the glass of the backing structure, the side of the
sandwich is covered with a thick layer of silicone and a plastic frame. For the
first prototype mirrors, this resulted in some small cracks in the thin glass
sheet. The addition of a short, L-shaped plastic border protecting the edge of
the thin glass sheet eliminated this problem, see Fig.~\ref{fig:edge_closeup}.
In May 2019, a larger production of these mirror types was started. As of that
date, several prototypes were produced and evaluated, the results of these
measurements are presented in the following section. Each mirror weighs 18\,kg,
about half compared to some of the first MAGIC mirrors that are still being used
(30--40\,kg), and similar compared to more recent mirrors (12--20\,kg).

\section{Measurements of Reflectivity and Point Spread Function}

\subsection{Laboratory}

Two 1\,m$^2$-sized prototypes according to the MAGIC standards were produced
(IDs~223 and~224). In the following months, four more (IDs~225--228) were
produced with smoother surface, and finally one mirror (ID~229) was produced
with improved edge protection, see closeup in Fig.~\ref{fig:edge_closeup}.

The evaluation of the reflectivity of the mirrors of these mirrors was performed
at the Max-Planck-Institute for Physics (MPP) in Munich~\cite{Scherpenberg}, the
focusing properties were checked at DESY Zeuthen at a test stand currently
evaluating mirrors for the Middle Size Telescopes (MST) of
CTA~\cite{Balkenkohl}. Before the mirror is released by MLT, several
measurements are done at the manufacturing site, which are used to cross check
the results obtained in the laboratories.

\begin{figure}[ht]
  \centering
  \includegraphics[width=.5\columnwidth]{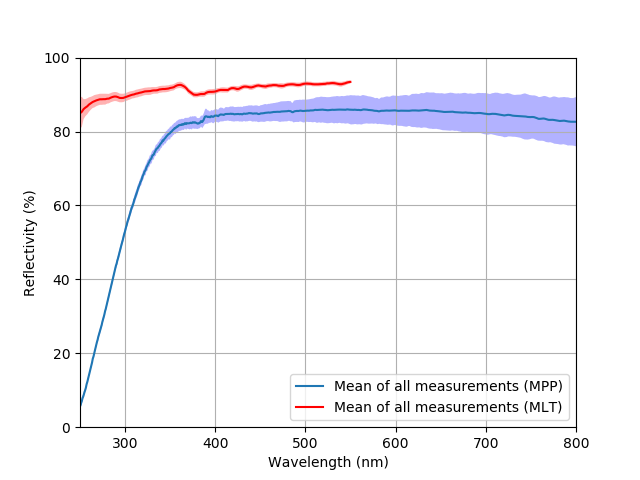}
  \caption{Mean surface reflectivity of the prototype back-coated mirrors. Shown
  is the average of the measurement on several points of the mirror. The MPP
  measurements are an average of mirrors 223--228. The measurements at MLT were
  done directly on the reflective aluminum surface, not on the thin glass sheet,
  which explains the higher reflectivity in the UV range.}
  \label{fig:mean_noncontact}
\end{figure}

At MPP, a non-contact device for measuring the thicknesses of a thin film and
its reflectivity was used to perform reflectivity measurements. It consists of a
deuterium-halogen light-source with glass fiber output, which would illuminate
from a short distance the sample under test. The light reflected from the sample
is collected by the fiber and analyzed with a spectrograph. For the measurement,
the 1\,m$^2$-sized samples were placed on a table and the sampling stage of the
device was rotated to allow placing the light output directly over the mirror,
see Fig.~\ref{fig:edge_closeup} for an image of the setup.  The averages of
those measurements of the mirrors are shown in Fig.~\ref{fig:mean_noncontact}.
Also shown are the reflectivity measurements done by MLT using an almost
identical non-contact device. The measurements are done directly on the
reflective aluminum surface, which explains the much higher reflectivity in the
UV range. Below 350\,nm, the reflectivity is reduced due to the transmission
properties of the selected ultra-thin glass sheet.

The setup at DESY used to measure the focal length and point spread function
(PSF) of the mirrors consists of a metal structure on which the mirror is fixed,
and a dark box containing the light source and electronics to measure the
reflected light. The mirror is placed at roughly twice its focal length with
respect to the dark box, this method to measure the PSF is called 2f setup.
Inside the dark box, an LED light source is installed on a movable carriage. A
collimator in front of the LED guides the outgoing light towards an exit hole in
the box through which it illuminates the mirror. Through another hole, the
reflected light enters and hits a reflective screen inside the dark box. Facing
the screen, a CCD camera is installed to take pictures of the reflected spot.

\begin{figure}
  \centering
  \begin{subfigure}{.48\textwidth}
    \centering
    \includegraphics[width=.65\columnwidth]{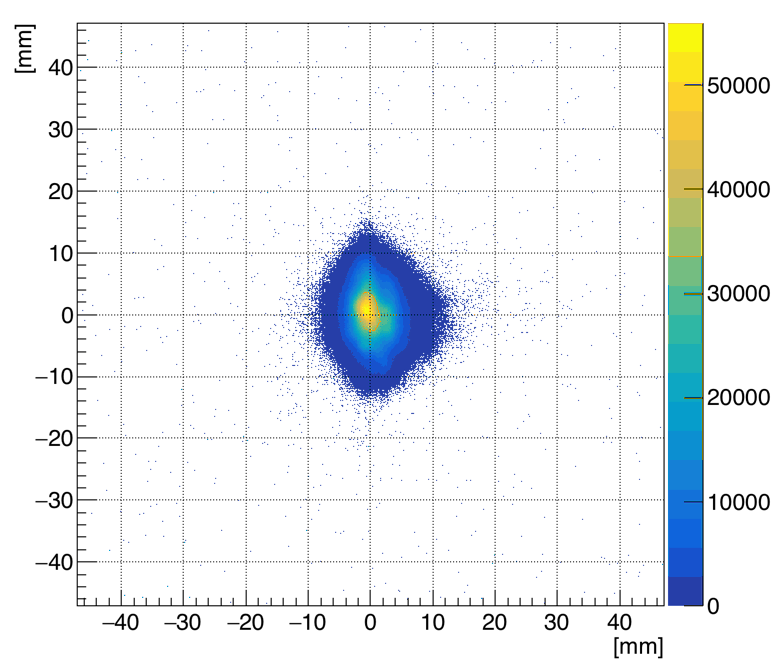}
  \end{subfigure}
  \hspace{1em}
  \begin{subfigure}{.48\textwidth}
    \centering
    \includegraphics[width=\columnwidth]{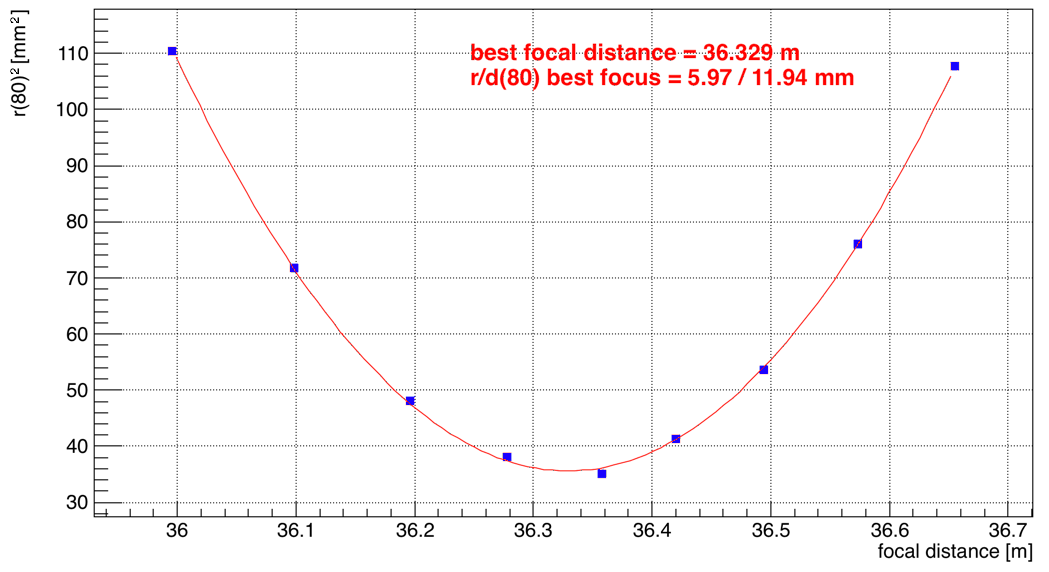}
  \end{subfigure}
  \caption{Left: Spot of mirror 227 around best focus. The 80\% containment
  $d_{80}$ is determined by summing up concentric rings around its center.
  Right: Focal scan, data points are fitted with a parabola. Its minimum yields
  the radius of curvature and the 80\% containment $d_{80}$.}
  \label{fig:spot_image_scan}
\end{figure}

The focal length and the spot size at the focus of a mirror are both determined
in a single measurement procedure. Several pictures of the spot are taken,
scanning the distance along the focal length by moving the carriage with light
source and camera inside the dark box. For each of these pictures, the PSF
diameter $d_{80}$ (radius $r_{80}$) of the circle which contains 80\% of the
reflected light is determined, see left panel of Fig.~\ref{fig:spot_image_scan}.
The right panel of Fig.~\ref{fig:spot_image_scan} shows $r^2_{80}$, which is
proportional to the area covered by the light spot, as a function of the
distance from the mirror. The data points are fitted with a parabola, whose
minimum yields the focal point $r_\mathrm{foc}$ and the PSF value of $d_{80}$ at
the best focus. The measurement at MLT is also done in a 2f setup, the light
source and camera are moved on a sled to change the distance to the mirror.
Several images are taken at varying distance to determine the best focus and
$d_{80}$ at that point. The determination of best focus is only done by hand
with an accuracy of about 50\,mm, no fit is performed. In addition, the surface
roughness and radius of the mirror $r$ are determined with a 3D scanner.

To see whether the focusing properties of the mirrors are constant after strong
temperature fluctuations, two mirrors (IDs~223 and~224) were put into a climate
chamber at DESY and underwent about five temperature cycles from $+30^\circ$C to
$-20^\circ$C during three days. After this, the measurements of PSF and
reflectivity were repeated. The results before and after the exposure to
temperature changes were consistent within the systematic uncertainties of the
measurement setup. The actual performance at extreme temperatures still needs to
be studied.

\begin{table}[ht]
  \centering
  \begin{tabular}{c|c|c|c|c|c|c|c|c|c|c|c}
       & Radius   & \multicolumn{2}{|c|}{Best focus}            & \multicolumn{2}{|c|}{PSF $d_{80}$ (mm)} & \multicolumn{6}{|c}{Reflectivity (\%)}    \\
    ID & $r$ (mm) & \multicolumn{2}{|c|}{$r_\mathrm{foc}$ (mm)} & \multicolumn{2}{|c|}{at best focus}     & \multicolumn{2}{|c|}{at 320\,nm} & \multicolumn{2}{|c|}{at 350\,nm} & \multicolumn{2}{|c}{at 400\,nm} \\
       & MLT      & MLT & DESY                                  & MLT & DESY                              & MLT & MPP                             & MLT & MPP                             & MLT & MPP                            \\
  \hline
  \hline
    223 & 36614 & 36600 & 36572 & 16.47 & 15.63 & --  & 67.3 & --   & 78.1 & --  & 84.1 \\
  \hline
    224 & 35873 & 36500 & 36511 & 14.43 & 11.27 & --  & 67.8 & --   & 79.0 & --  & 83.8 \\
  \hline
    225 & 35839 & 36100 & 36197 & 16.50 & 14.28 & 91.8 & 67.9 & 92.7 & 79.9 & 91.6 & 84.0 \\
  \hline
    226 & 35934 & 36300 & 36400 & 16.99 & 11.36 & 91.1 & 68.3 & 92.0 & 79.2 & 90.8 & 84.1 \\
  \hline
    227 & 36004 & 36300 & 36329 & 14.95 & 11.94 & 90.0 & 68.4 & 91.2 & 80.3 & 90.3 & 84.6 \\
  \hline
    228 & 36013 & 36800 & 36742 & 12.70 & 12.70 & 89.2 & 69.2 & 90.4 & 80.7 & 89.8 & 85.3 \\
  \hline
    229 & 35983 & 36650 & --    & 13.40 & --    & 91.7 & --   & 92.6 & --   & 91.7 & --   \\
  \end{tabular}
  \caption{Results of the laboratory measurements of the seven mirror prototypes. The MLT reflectivity measurements were done on the reflective side, the MPP measurements
  through the thin glass sheet. Mirror~229 was not yet measured at MPP or DESY.}
  \label{tab:meas_results}
\end{table}

In Tab.~\ref{tab:meas_results}, the measurements done at MPP, DESY and MLT are
summarized. For mirrors 223 and 224, no measurements from the manufacturing site
at MLT are available, mirror~229 has not been measured at MPP or DESY. The
results of the best focus measurements from DESY and MLT agree within 100\,mm,
which is explained by the rough distance scan at MLT. The 3D measurements of the
radius show larger differences to the best focus values, almost 800\,mm for
mirror~228, and systematically smaller radii. The 3D measurements are done with
the mirror lying flat on a table, while for the 2f setup the mirror is hanging
vertically. The mirror is deforming and the measured radius is different. The
surface roughness determined by the 3D measurement is between 4 and
12\,\,\textmu m (RMS) with peak-to-valley values between 20 and 85\,\textmu m.

\subsection{In situ}

\begin{figure}[ht]
  \centering
  \includegraphics[width=.7\columnwidth]{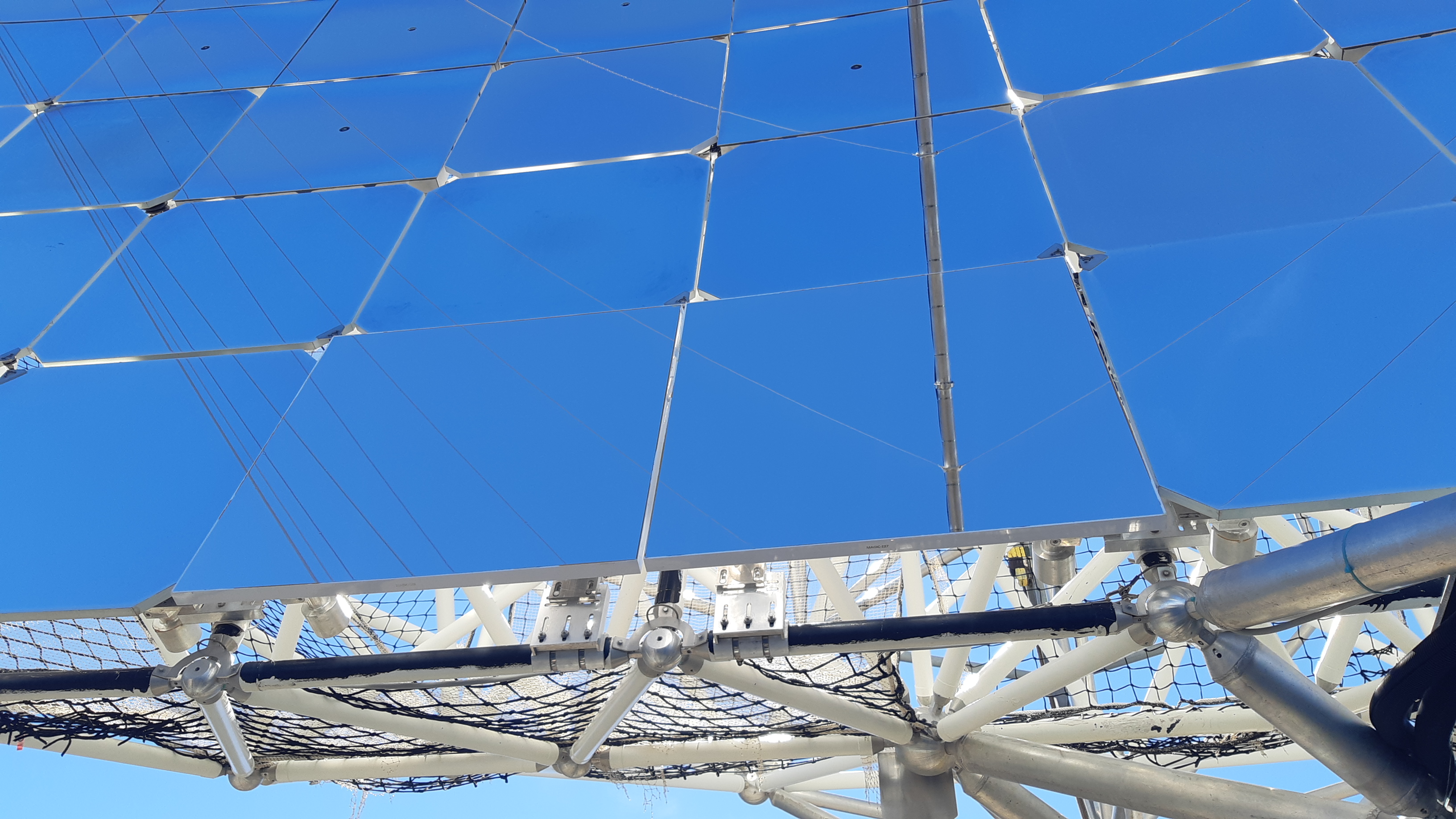}
  \caption{Two mirrors, IDs 227 and 228, were installed in the MAGIC-II
  telescope in October 2018 to evaluate their performance and long-term behavior
  under actual on-site condition.}
  \label{fig:installed}
\end{figure}

In October 2018, prototype mirrors 227 and 228 were installed in the MAGIC-II
reflector. They have a curvature radius around 36\,m, which is the maximum
radius of the mirrors on the outside of the dish, the two mirrors were installed
at the bottom edge, see Fig.~\ref{fig:installed}. This has the additional
benefit of making visual inspection easier when the telescope is parked. In the
9~months since the installation, no change in the surface or the edges was
observed.

The PSF and reflectivity of the mirrors is checked periodically by tracking a
bright star, focusing the reflected light on a square target in front of the PMT
camera, focused at infinity. An image is taken using a CCD camera mounted in the
dish center, containing the direct light from the star and the reflected light
on the target~\cite{absolutereflectance}. The $d_{80}$ of the PSF is found by
integrating the light of the reflected spot and determining the 80\%
containment. The reflectivity is determined by dividing the integrated light of
the reflected spot by that of the direct star light, applying corrections for
the reflectivity of the target, mirror area, opening angle of the CCD camera and
other geometrical factors. While the total reflectivity of the mirror is an
important parameter, the reflectivity within $d_{80}$ is an indicator of the
focused reflectivity. The light reflected far from the central spot does not
contribute in the central pixel and is a source of background in other pixels of
the camera.

\begin{figure}
  \centering
  \begin{subfigure}{.45\textwidth}
    \centering
    \includegraphics[width=\columnwidth]{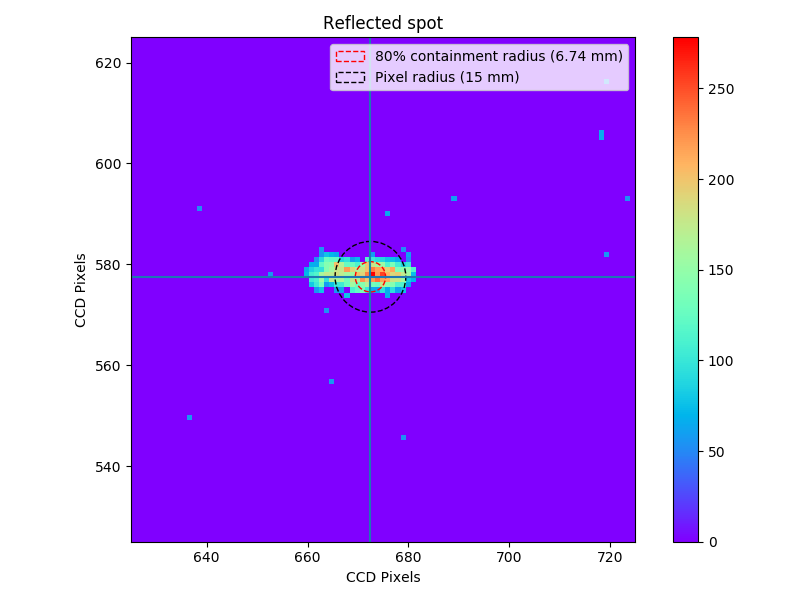}
  \end{subfigure}
  \begin{subfigure}{.45\textwidth}
    \centering
    \includegraphics[width=\columnwidth]{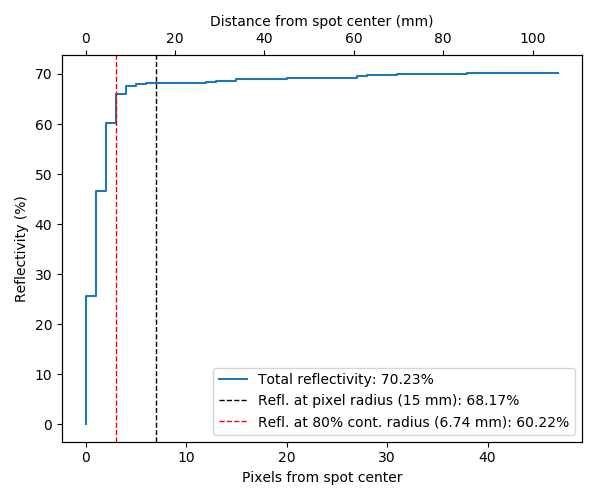}
  \end{subfigure}
  \caption{Point Spread Function and reflectivity of mirror 227.}
  \label{fig:psf_refl_227}
  \begin{subfigure}{.45\textwidth}
    \centering
    \includegraphics[width=\columnwidth]{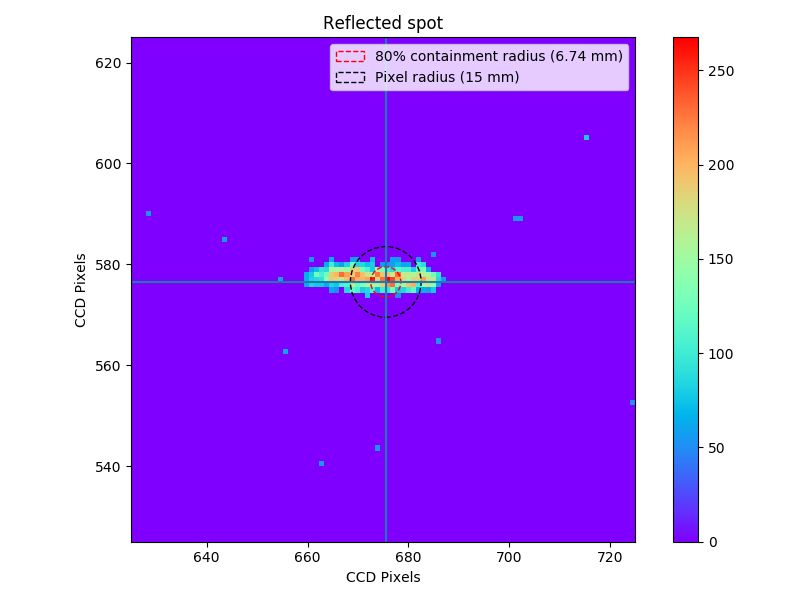}
  \end{subfigure}
  \begin{subfigure}{.45\textwidth}
    \centering
    \includegraphics[width=\columnwidth]{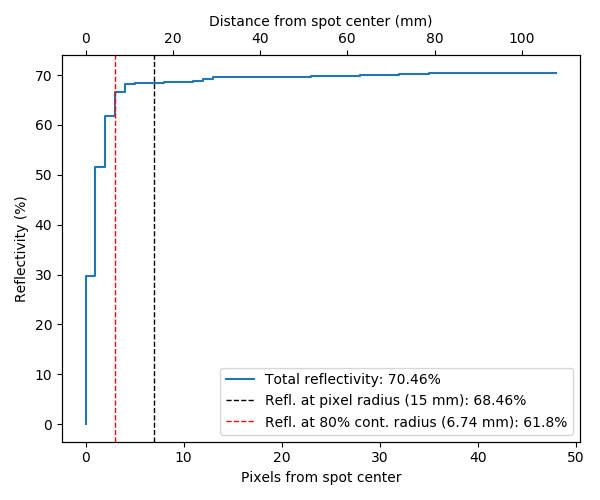}
  \end{subfigure}
  \caption{Point Spread Function and reflectivity of mirror 228.}
  \label{fig:psf_refl_228}
\end{figure}

In Fig.~\ref{fig:psf_refl_227} and~\ref{fig:psf_refl_228} the results are shown
for the two prototypes. The reflected spot is very elongated, this aberration is
a result of the placement of the mirrors at the edge of the reflective surface.
The mirrors are spherical but the ideal shape of the reflector is parabolic, at
large distances to the center the difference in curvature between the mirror and
the ideal shape is different along the axes of the mirror and the spot is
distorted in one direction. To correct for this, the 80\% containment radius was
evaluated along the projection of the reflected light along the axis that is not
distorted. The $d_{80}$ values of 13.5\,mm for both mirrors, the focused
reflectivity of 60.2\% and 61.8\%, and the total reflectivity of 71.5\% and
69.5\% are very good.

\section{Outlook}

In conclusion, the novel back-coated mirror samples show excellent focusing
properties, high reflectivity, and they are stable under short-term temperature
variations. These mirrors are expected to have a lifetime of a few decades, on
the order of the lifetime of a telescope, and are much easier to maintain than
the mirrors that are currently being used in Cherenkov telescopes. Two of these
back-coated mirror panels were installed on MAGIC in October 2018 to test their
longterm stability, and 29~more will be installed in 2019 to replace part of the
10--15~year old original mirrors that show strong signs of degradation. This
will allow to evaluate the performance of a significant sample of these novel
mirrors, which will provide an important input for the CTA collaboration.

\paragraph{Acknowledgements}

\href{https://magic.mpp.mpg.de/acknowledgments&#95;ICRC2019/}{https://magic.mpp.mpg.de/acknowledgments\_ICRC2019/}

\end{document}